\documentclass[conference]{IEEEtran}
\IEEEoverridecommandlockouts
\usepackage{cite}
\usepackage{amsmath,amssymb,amsfonts}
\usepackage{algorithmic}
\usepackage{graphicx}
\usepackage{textcomp}
\usepackage{xcolor}
\def\BibTeX{{\rm B\kern-.05em{\sc i\kern-.025em b}\kern-.08emgpffl0813-

    T\kern-.1667em\lower.7ex\hbox{E}\kern-.125emX}}

\newcommand{\norm}[1]{\left\lVert #1 \right\rVert}
\usepackage{amsmath,bm}
\usepackage{optidef}
\usepackage{algorithm}
\usepackage{stackengine}
\def\delequal{\mathrel{\ensurestackMath{\stackon[1pt]{=}{\scriptscriptstyle\Delta}}}}
\usepackage{tabu}
\usepackage{multirow}

\usepackage{geometry}
\usepackage{amsthm}
\theoremstyle{remark}

\usepackage{easybmat}
\usepackage{colortbl}
\definecolor{Gray}{gray}{0.92}
\usepackage{hhline}
\usepackage{stackengine}

\makeatletter
\newcommand{\vast}{\bBigg@{3}}
\newcommand{\Vast}{\bBigg@{4}}
\makeatother

\makeatletter
\newcommand{\vvast}{\bBigg@{4}}
\newcommand{\vVast}{\bBigg@{4}}
\makeatother

\def\capmystringaux#1#2\relax{\uppercase{#1}\lowercase{#2}}
    \IEEEoverridecommandlockouts
\IEEEpubid{\makebox[\columnwidth]{978-1-7281-4490-0/20/\$31.00 \copyright~2020~IEEE }
\hspace{\columnsep}\makebox[\columnwidth]{ }}
\begin{document}

\title{ Rate-Splitting Multiple Access for Multi-Antenna Broadcast Channel with Imperfect CSIT and CSIR% Rate-Splitting Multiple Access for Downlink Multi-User Multi-Antenna Communications with Imperfect CSIT and CSIR
 \\

}

\author{\IEEEauthorblockN{Jihye An$^\dagger$, Onur Dizdar$^\ddagger$, Bruno Clerckx$^\ddagger$, and Wonjae Shin$^\dagger$}
\IEEEauthorblockA{{$^\dagger$Department of Electronics Engineering, Pusan National University, Busan, Korea}\\
{ $^\ddagger$Department of Electrical and Electronic Engineering, Imperial College London, United Kingdom }\\
Email: $^\dagger$\{jh\_an, wjshin\}@pusan.ac.kr, $^\ddagger$\{o.dizdar, b.clerckx\}@imperial.ac.uk}
}

\maketitle

\begin{abstract}
%This paper studies an optimization of precoding vectors with aim to maximize sum-rate in downlink mutiuser multiple-input single-output (MU-MISO) broadcast channels where a rate-splitting (RS) approach is adapted. %which is unusual transmission scheme.
Rate-splitting multiple access (RSMA) has appeared as a powerful transmission and multiple access strategy for multi-user multi-antenna communications. 
Uniquely, this paper studies the optimization of the sum-rate of RSMA with imperfect channel state information (CSI) at the transmitter (CSIT) and the receivers (CSIR).
Robustness of the RSMA approach against imperfect CSIT has been investigated in the previous studies while there has been no consideration for the effects of imperfect CSIR.
This motivates us to develop a  robust design relying on RSMA in the presence of both imperfect CSIT and CSIR. 
Since the optimization problem for the design of RSMA precoder and power allocations to maximize the sum-rate  is non-convex, it is hard to solve directly.
To tackle the non-convexity, we propose a novel alternating optimization algorithm based on semidefinite relaxation (SDR) and concave-convex procedure (CCCP) techniques.
%
%The proposed algorithm can be extended to conventional methods with slight modification and its results are used to compare the results of the proposed RS.
%
%The proposed algorithm is extended to conventional scheme by modifying.
%By comparing simulation results of the RS with the conventional methods,
%In the simulation results comparison,
By comparing simulation results with conventional methods, it turns out that RSMA is quite robust to imperfect CSIR and CSIT, thereby improving the sum-rate performance.
% Also, it is numerically showed that superiority of the optimized the percoding vector with proposed algorithm compared to the fixed precoding approaches.

%the robustness of RS aginst imperfect CSIT is already shown in previous studies. 

%to obtain a sub-optimal solution. 

%this paper deals with  for sum-rate maximization.
%RS is a scheme splitting messages into common and private parts. The common message is decoded by all users and private message is decoded by only corresponding user. 
%We assume that the receivers imperfectly know channel state information (CSI) and the transmitter obtains the imperfect CSI from the receiver via a lossless feedback link.
%Under imperfect CSI, the achievable rate at receiver is determined by using generalized mutual information (GMI). We formulate sum-rate maximization problem that is difficult to solve its non-convexity.
%Thus, we propose a novel algorithm based on semidefinite relaxation (SDR) and concave-convex procedure (CCCP) techniques to obtain a sub-optimal solution. % In non-convex quadratically constrained quadratic program (QCQP), SDR can obtain a solution close to the optimal solution. By using CCCP, a local optimal solution can be obtained as a solution of difference of convex functions (DC) problem that is non-convex.
%The sub-optimal solution and local solution can be obtain by using SDR and CCCP respectively. 
%Simulation result shows that the proposed algorithm improves the sum-rate compared to existing conventional methods.
\end{abstract}


\section{Introduction}
%Recently, the data traffic is exponentially increasing due to the increase in the number of mobile devices and users with development of mobile devices and communication technology. As a result, the interest in high data rate and efficient use of resources is increasing. Also, It occurs multiple input multiple output (MIMO) downlink system suffering from significant interference.
Due to the increase in data traffic and the number of communicating devices, there is an increasing need to design efficient communication strategies to boost the data rate, spectral efficiency and manage the interference. 
To that end, multi-antenna/multiple-input multi-output (MIMO) processing is a key technology.
 To deal with the interference problem in multi-user multi-antenna systems, the perfect channel state information (CSI) at receiver (CSIR) and transmitter (CSIT) are essential.
 However, it is difficult to obtain accurate CSI due to quantization error, channel mobility, and estimation error. 
 Even with the ideal assumption of perfect CSIR, it is questionable whether a base station (BS) can obtain accurate CSIT.

Rate-splitting multiple access (RSMA) has recently emerged and has been found to have multiple advantages over conventional multiple access methods in terms of robustness against imperfect CSIT \cite{RS:rob}, and spectral and energy efficiencies \cite{RS:mag}, \cite{RS:eff}. 
 The key feature of RSMA is the split of the messages into common and private parts. The common parts are encoded in a common stream that can be decoded by multiple users.
 On the other hand, each of the private messages is encoded in a private stream which is decoded by its respective receiver. 
 Each receiver then decodes the common stream, retrieves its intended common part, then removed the common stream from the received signal using successive interference cancellation (SIC). %
 After removing the common stream, each receiver can decode its intended private stream by treating the remaining private streams as interferences.
 From the common stream and the private stream, each receiver can reconstruct the original message. 
 The flexibility of RSMA lies in adjusting the content and power allocated to the common and private streams, so as to partially decode interference and partially treat interference as noise \cite{RS:Bridging}.
 Such flexibility leads to more robustness and performance enhancements in various network and propagation conditions \cite{RS:Bridging}.

 %The common message can be decoded multiple users. Otherwise, the private message can be decoded only corresponding user. Therefore, after decoding the common signal, the common signal can be removed by using successive interference cancellation (SIC). Thus, when decoding private part, there is only interference from the other private signal not common signal i.e., part of the interference is decoded and the other part is treated as noise. This special method for interference management makes RS can provide benefit in performance.

In \cite{RS:uni},\cite{RS:uni_multi}, it shown that RSMA can outperform conventional approaches in terms of rate maximization under perfect CSI assumption. Especially in \cite{RS:uni}, it is shown that RSMA unifies other four strategies (i.e, non-orthogonal multiple access (NOMA), space-division multiple access (SDMA), orthogonal multiple access (OMA) and multicasting) and outperforms them in a two user multiple-input single-output (MISO) broadcast channel (BC) channel. 

In case of imperfect CSIT and perfect CSIR scenario, the BS is unable to calculate the achievable rates at the receivers accurately. Thus, the BS should adjust precoding vector and power allocation by using estimated channel and error information. In \cite{RS:imperfect}, sample average approximation combined with a weighted minimum mean square error (WMMSE) algorithm is used for sum-rate maximization in RSMA by generating channel error ensembles.
In \cite{RS:rob}, max-min fairness optimization using the worst case rate and WMMSE is proposed with bounded channel error. Both studies show robust transmission of RSMA in multi-user (MU) MISO compared to conventional approaches. 

 In this paper, we consider both imperfect CSIR and CSIT under the assumption that the BS obtains CSI from the receiver through lossless channel feedbacks.
 This is the first paper studying the design and optimization of RSMA with both imperfect CSIT and CSIR.
 We formulate the sum-rate maximization problem in RSMA based MU-MISO system. 
 For converting non-convex problem to convex problem, the algorithm using the two methods semidefinite relaxation (SDR) and concave-convex procedure (CCCP) is proposed. 
 By the proposed algorithm, we jointly optimize precoding vectors and power allocation.
 In simulation, we show the performance gains of the proposed RSMA over existing techniques.
 
The reminder of this paper is organized as follows. In section II, system model and achievable rate in imperfect CSI are described. In section III, the optimization problem for maximizing the sum-rate is formulated and joint precoding vector and power allocation optimization is conducted for sum-rate optimization by the proposed algorithm based on SDR and CCCP. Simulation result are provided in section IV. The paper is concluded in V.

%The achievable degrees-of-freedom (DoF) regions in RS MIMO is investigated in \cite{RS:MIMO:DOF}.  
%The assumption of perfect CSIR and CSIT seems to be idealistic scenarios as the CSI is usually imperfect in practical situation. The performance of RS in perfect CSIR and imferpect CSIT is treat much. Otherwise, imperfect CSIR scenario hasn't been covered much yet.
%
%However, most of the works on RS consider
%

\subsection{Notaion}
Standard letter indicates scalar, lower case boldface letter denotes vector, and upper case boldface letter denotes matrix.
Notation $\mathbf{A}\succeq \mathbf{B}$ indicates that matrix $\mathbf{A}- \mathbf{B}$ is positive semidefinite matrix.
Superscript $(\cdot)^{H}$ denotes hermitian (conjugate transpose).
Trace of matrix $\mathbf{A}$ is denoted by $\mathrm{tr}(\mathbf{A})$ and 
rank of matrix $\mathbf{A}$ is denoted by $\mathrm{rank}(\mathbf{A})$.
Notations of $|\cdot |$, $||\cdot ||$, and $ \mathbb{E}[\cdot]$ refer to the absolute value, Euclidean norm, and expectation operation, respectively. A matrix $\mathbf{I}_n$ denotes a $n$ by $n$ identity matrix.
\begin{figure}[t]
\includegraphics[width=1\linewidth]{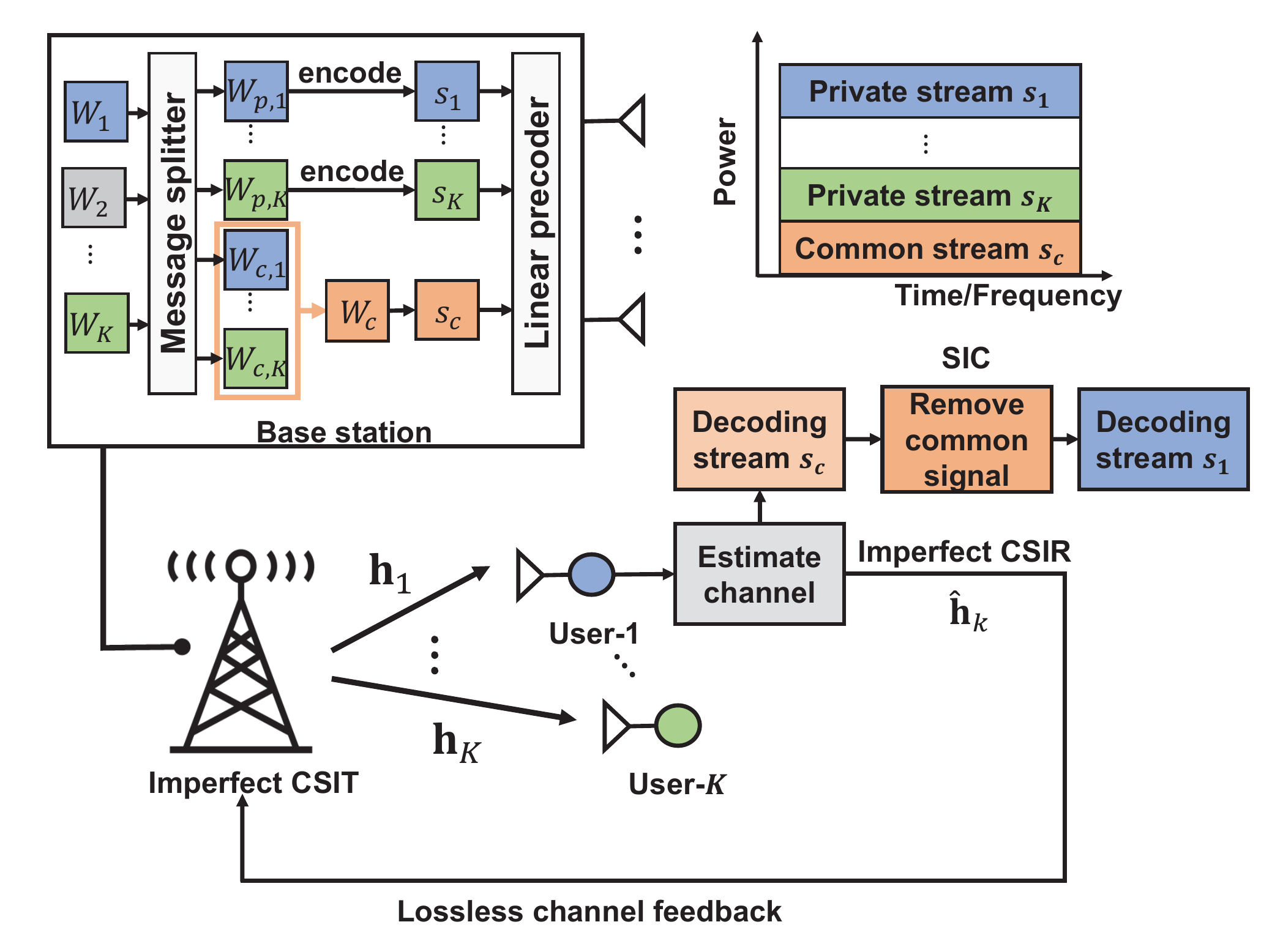}
\centering
\caption{System architecture of rate-splitting multiple access with imperfect CSIR and CSIT in MU-MISO BC}
\label{System}
\end{figure}
\section{System Model}
\subsection{Rate-Splitting Multiple Access Based System
%/ signal model with rate-splitting
}
We consider a single cell MU-MISO system operating in downlink where the BS equipped with $N_{t}$ antennas serves $K$ single antenna users.
As shown in Fig. \ref{System}, the main idea of RSMA is to split a message for user-$k$ $W_{k}$, $k=1,\dots, K$ 
%$k \in \mathcal{K}\delequal\{1, \dots, K\}$, 
into common and private parts, i.e. $W_{k}=\{W_{p,k},W_{c,k}\}$. 
The common part can be decoded by all users and the private part can be decoded by only the corresponding user. All common parts of each user message are combined into one common message $W_{c}$, i.e. $W_{c}=\{ W_{c,1},\dots,W_{c,K} \}$. The common message $W_{c}$ is encoded into the common stream $s_{c}$ by using a codebook known to all users and each private message $W_{p,k}$ is encoded into the private stream $s_{k}$ by using a codebook known to only the intended receiver.
 Each stream is assumed to be independent zero mean unit variance Gaussian random variable, i.e. $s_{i}\sim \mathcal{CN}(0,1),~i \in \mathcal{I}\delequal\{c, 1,\dots, K\} .$  These $K+1$ streams are linearly precoded by using precoding vector $\mathbf{p}_{i} \in \mathbb{C}^{N_{t}\times1},~i\in \mathcal{I}$.
 The transmitted signal at the BS is expressed as
 \begin{align}
\mathbf{x}&=\mathbf{p}_\mathrm{c}s_\mathrm{c}+\sum_{k=1}^{K}\mathbf{p}_{k}s_{k},\label{tranmit_signal}
\end{align}
where a transmitted signal power constraint with a total power $P_t$ is 
\begin{align}
\sum\limits_{\substack{i\in\mathcal{I}}}\norm{\mathbf{p}_{i}}^2\leq P_{t}.
\end{align} 
%
%The squared norm of precoding vector indicates the power allocated to corresponding stream.  
%
We refer to $\mathbf{h}_{k}\in \mathbb{C}^{N_{t}\times1}$ as a downlink channel vector from the BS to user-$k$ and a received signal at user-$k$ is denoted by
\begin{equation}
y_{k}=\mathbf{h}_{k}^{H}\boldsymbol{\mathrm{x}}+n, ~k=1, \dots, K,\label{received_signal}
\end{equation}
where $n\sim \mathcal{CN}(0,\sigma_{{n}}^2)$ is additive white Gaussian noise (AWGN). Since the common stream can be decoded by all users, users can remove the common stream by SIC. Thus, users decode the private stream after SIC. 
When decoding the common stream, all private streams are treated as interference. When decoding a private stream, only other private streams are treated as interference, provided that the common stream is completely removed. 
Each user reconstructs the original message after retrieving the part of its message encoded in the common stream and the part encoded in the private stream.

\subsection{Assumption on Channel State Information}
We assume that users cannot accurately estimate the channel vector, i.e. imperfect CSIR.
The channel model is given by
\begin{equation}
\mathbf{h}_{k}=\hat{\mathbf{h}}_{k}+\mathbf{e}_{k},\label{channel}
\end{equation}
where $\hat{\mathbf{h}}_{k}$ is an estimated channel and $\mathbf{e}_{k}\sim \mathcal{CN} (0,\mathbf{\Phi}_k)$ is a channel error. 
Also, we assume that the BS has the same CSI with the users because of lossless channel feedback. 
Thus, all users and the BS know the expectation of the channel, $\mathbb{E}[\mathbf{h}_k]=\hat{\mathbf{h}}_{k}$, and the covariance of the channel, $\mathbb{E}[(\mathbf{h}_k-\mathbb{E}[\mathbf{h}_k])(\mathbf{h}_k-\mathbb{E}[\mathbf{h}_k])^H]=\mathbf{\Phi}_k$. 
In this paper, it is assumed that the covariance matrix of channel error $\mathbf{e}_{k}$ is  $\mathbf{\Phi}_k=\sigma _{\mathrm{e},k}^2\mathbf{I}$. In  other words, the channel error is assumed as a vector of independent and identically distributed (i.i.d) random variables.
\subsection{Achievable Rate}
It is difficult to determine an explicit achievable rate under the imperfect CSIR assumption, since the users do not know the actual channel. Thus, the concept of generalized mutual information (GMI) is used in order to characterize the achievable rate at a user with imperfect CSI \cite{GMI:1},\cite{GMI:2}. We first introduce a general form of GMI by considering a point to point case for simplicity. When the input has Gaussian distribution $x\sim \mathcal{CN} (0,\epsilon_x)$, the output signal is expressed by
\begin{align}
y=hx+n
\label{y:GMI}
\end{align}
where $h$ is the fading channel and $n\sim \mathcal{CN} (0,N)$ is the noise. When knowing expectation and variance of channel, $h$ can be broken into $\hat{h}$ and $e$, i.e. ${h}=\hat{h}+e$ where $\mathbb{E}[h]=\hat{h}$ and $\mathbb{E}[e]=0$. We can intuitively consider $\hat{h}$ as an estimate of the channel and $e$ as a channel error having zero mean with variance $\sigma_h^2$. GMI is defined by
\begin{align}
{I_\mathrm{GMI}}=\log_2\left(1+\frac{|\hat{h}|^2 \epsilon_x}{\mathbb{E}[|e|^2]\epsilon_x+N}\right),
\label{I:GMI}
\end{align}
where $\mathbb{E}[|e|^2]=\sigma^2_{h}$. 
%Also, it can work as lower bound of achievable rate at receiver with perfect CSIR \cite{GMI:ex1}.
In the case of imperfect CSIR, GMI corresponds to an achievable rate when a user uses a nearest neighbor decoder and the input is Gaussian distribution \cite{GMI:3}. %\cite{GMI:4}.
By using this property, we apply GMI to RSMA based system and derive the achievable rate under imperfect CSI. 

In RSMA approach, a user first decodes the common stream and then decodes the corresponding private stream after SIC. By this feature, the rate of the private stream is derived using the received signal after SIC.
The received signal at user-$k$ in (\ref{received_signal}) is rewritten as
\begin{align}
y_{k} &=\hat{\mathbf{h}}_{k}^{H}\mathbf{x}+\mathbf{e}_{k}^{H}{\mathbf{x}}+n, ~k=1,\dots ,K \label{yk2}\\
 &=\hat{\mathbf{h}}_{k}^{H}\mathbf{p}_\mathrm{c}s_\mathrm{c}
 +\mathbf{e}_{k}^{H}\mathbf{p}_\mathrm{c}s_\mathrm{c} \nonumber
 \\ & \,\,\,\,\,\,\,\,\,\,\,\,\,\,\,+\sum_{j=1}^{K}(\hat{\mathbf{h}}_{k}^{H}\mathbf{p}_{j}s_{j}+\mathbf{e}_{k}^{H}\mathbf{p}_{j}s_{j})+n.\label{yk3}
\end{align}
Considering the common stream, the signal received from user-$k$ in (\ref{yk3}) can be re-expressed in the form of (\ref{y:GMI}) as
% When considering rate of common signal, the received signal can be re-expressed by
 %
 \begin{align}
y_{k} &=\hat{h}_{k,c}s_\mathrm{c}
 +e_{k,c} s_\mathrm{c}
 +z_\mathrm{c}\\
 &={h}_{k,c}s_\mathrm{c} +z_\mathrm{c}.
\end{align}
where ${h}_{k,c}=\hat{h}_{k,c}+e_{k,c}$, 
$\hat{h}_{k,c}=\hat{\mathbf{h}}_{k}^{H}\mathbf{p}_\mathrm{c}$, 
$e_{k,c}=\mathbf{e}_{k}^{H}\mathbf{p}_\mathrm{c}$, and $z_\mathrm{c}=\sum_{j=1}^{K}(\hat{\mathbf{h}}_{k}^{H}\mathbf{p}_{j}s_{j}+\mathbf{e}_{k}^{H}\mathbf{p}_{j}s_{j})+n$. 
Due to the independence between each stream and the noise, the expectation and variance of each component are derived as follows:
 \begin{align} 
 &\mathbb{E}[{h}_{k,c}]=\hat{h}_{k,c}=\hat{\mathbf{h}}_{k}^{H}\mathbf{p}_\mathrm{c},\label{aver1}\\
&\mathbb{E}[e_{k,c}]=0,
~\mathbb{E}[z_\mathrm{c}]=0,\\
&\mathbb{E}[|e_{k,c}|^2]=\mathbb{E}[|\mathbf{e}_{k}^{H}\mathbf{p}_\mathrm{c}|^2],\label{aver2}\\
&\mathbb{E}[|z_\mathrm{c}|^2]=\sum\limits_{j=1}^{K} ({|\hat{\mathbf{h}}{}^{H}_{k}\mathbf{p}_{j}|}{}^{2}+\mathbb{E}[|\mathbf{e}_{k}^{H}\mathbf{p}_{j}|^2])+\sigma_n^2.\label{aver5}
\end{align}
Substituting the values in (\ref{aver1}), (\ref{aver2}) and (\ref{aver5}) into (\ref{I:GMI}), the GMI for the common stream under imperfect CSIT can be obtained as 
%When applying these values to GMI in (\ref{I:GMI}), GMI which is the achievable rate of common stream in imperfect CSI can be obtained with

\begin{align}
R_{c,k}&=\log_{2}  \vast(1+\frac{|\hat{\bm{\mathrm{h}}}{}^{H}_{k}\mathbf{p}_\mathrm{c}|^{2}}{\sum\limits_{j=1}^{K} {|\hat{\mathbf{h}}{}^{H}_{k}\mathbf{p}_{j}|}{}^{2}+\sum\limits_{\substack{j\in \mathcal{I}}}{\mathbb{E}[|\mathbf{e}^{H}_{k}\mathbf{p}_{j}|^{2}]}+\sigma _{{n}}^2}\vast),
\label{Rck1}
\end{align}
where $\mathbb{E}[|\mathbf{e}^{H}_{k}\mathbf{p}_{j}|^{2}]
={\mathbf{p}_{j}^\mathrm{H}\mathbf{\Phi}_k\mathbf{p}_{j}}$, due to $\mathbb{E}[\mathbf{e}_{k}{\mathbf{e}_{k}}^H]=\mathbf{\Phi}_k$. Note that $|\mathbf{e}^{H}_{k}\mathbf{p}_{c}|^{2}$ is associated with not only the desired stream but also the channel error. Thus this term is considered as interference when decoding the desired stream, since users do not have any information on the channel error. The same phenomenon occurs when decoding the private streams. 

When operating with perfect CSIR, the common stream can be removed perfectly by SIC.
%One of the characteristics of RSMA is that the common stream can be removed by SIC when CSIR is perfect. 
%However, CSIR is imperfect, users cannot perfectly remove the common stream because users do not know the accurate channel. 
%
However, the common stream cannot be removed perfectly under imperfect CSIR, since the users do not have accurate information the actual channel.
Thus, the part of the common stream associated with channel error still remains after SIC. The received signal after SIC with imperfect CSIR is expressed by
  \begin{align}
y_{k,\mathrm{SIC}} &=y_{k}-\hat{\mathbf{h}}_{k}^{H}\mathbf{p}_\mathrm{c}s_\mathrm{c}\\
 &=\mathbf{e}_{k}^{H}\mathbf{p}_\mathrm{c}s_\mathrm{c} +\sum_{j=1}^{K}(\hat{\mathbf{h}}_{k}^{H}\mathbf{p}_{j}s_{j}+\mathbf{e}_{k}^{H}\mathbf{p}_{j}s_{j})+n.
\end{align}
The rate of the private stream can be obtained in a similar manner as the common part. The received signal after SIC is rewritten as 
\begin{align}
y_{k,\mathrm{SIC}} &=\hat{h}_{k}s_{k}
 +e_{k} s_{k}
 +z_{k}\\
 &={h}_{}s_{k} +z_{k}.
\end{align}
where ${h}_{k}=\hat{h}_{k}+e_{k}$, 
$\hat{h}_{k}=\hat{\mathbf{h}}_{k}^{H}\mathbf{p}_{k}$, 
$e_{k}=\mathbf{e}_{k}^{H}\mathbf{p}_{k}$, and $z_{k}=\mathbf{e}_{k}^{H}\mathbf{p}_{c}s_\mathrm{c}+\sum_{\substack{j=1\\j\neq k}}^{K}(\hat{\mathbf{h}}_{k}^{H}\mathbf{p}_{j}s_{j}+\mathbf{e}_{k}^{H}\mathbf{p}_{j}s_{j})+n$.
Thus, the achievable rate of private stream for user-$k$  with imperfect CSIR is determined as

\begin{align}
R_{k}
&=\log_{2}\vast(1+\frac{{|\mathbf{\hat{h}}{}^{H}_{k}\mathbf{p}_{k}|}{}^{2}}{\sum\limits_{\substack{j=1\\j\neq k}}^{K} |\hat{\mathbf{h}}{}^{H}_{k}\mathbf{p}_{j}|^{2}\!\!+\!\!\sum\limits_{j\in \mathcal{I}}{\mathbb{E}[|\mathbf{e}^{H}_{k}\mathbf{p}_{j}|^{2}]}+\sigma _{{n}}^2}\vast).
\label{Rk2}
\end{align}
%Using these rate equation, we formulate and optimize the sum-rate.

%
%
\section{ Sum-rate Maximization with imperfect CSI  }
In this section, a sum-rate maximization problem is formulated. We transform the optimization problem and propose a novel algorithm for solving the optimization problem that is non-convex.
\subsection{Problem Formulation}
Our objective is to optimize precoding vectors consisting of power and direction for maximizing the sum-rate. The sum-rate is expressed as \begin{align}R_{s}=R_{c}+\sum\limits_{k=1}^{K}R_{k},
\end{align}
in which the common rate  $R_{c}\delequal \min_{k} R_{c,k}, ~k=1,\dots,K$ because the common stream is decoded by all users.
The optimization problem for sum-rate maximization is formulated as:
%P1
\begin{maxi!}|s|[0]
	{\substack{\mathbf{p}_{i},\forall i }}{R_{c}+\sum\limits_{k=1}^{K}R_{k}}
	{\label{P1}}{\bm{\mathsf{(P1):}} \nonumber}
	\addConstraint{R_{c,k} \geq  R_{c} \label{P1:const1}}
	\addConstraint{\sum\limits_{\substack{\forall i \in \mathcal{I} }}\norm{\mathbf{p}_{i}}^2\leq P_{t}. \label{P1:const2}}
%	\addConstraint{p_n}{\geq 0, \, \forall n \in \mathcal{N}, \label{P1:const1}}
\end{maxi!}
We first simplify expression of the objective function of $\bm{\mathsf{(P1)}}$ using a stacking method introduced in \cite{GMI:ex1}.
First, we equivalently transform the expressions of the rate (\ref{Rck1}) and (\ref{Rk2}) as
\begin{align}
R_{c,k}&=\log_{2}\vvast(\frac{\sum\limits_{\substack{j\in \mathcal{I}}}{\mathbf{p}_{j}^{H}(\hat{\mathbf{h}}_{k}\hat{\mathbf{h}}{}^{H}_{k}+\mathbf{\Phi}_k)\mathbf{p}_{j}}+\sigma _{{n}}^2}{\sum\limits_{j=1}^{K} {\mathbf{p}_{j}^{H}\hat{\mathbf{h}}_{k}\hat{\mathbf{h}}{}^{H}_{k}\mathbf{p}_{j}}+\sum\limits_{\substack{j\in \mathcal{I}}}{\mathbf{p}_{j}^{H}\mathbf{\Phi}_k\mathbf{p}_{j}}+ \sigma _{{n}}^2}\vvast)\label{Rck1_2}
\end{align}
and
\begin{align}
R_{k}=\log_{2}\vvast(\frac{\sum\limits_{\substack{j=1}}^{K} \mathbf{p}_{j}^{H}\hat{\mathbf{h}}_{k}\hat{\mathbf{h}}{}^{H}_{k}\mathbf{p}_{j}\!\!+\!\!\sum\limits_{j\in \mathcal{I}}{\mathbf{p}_{j}^{H}\mathbf{\Phi}_k\mathbf{p}_{j}}+\sigma _{{n}}^2}{\sum\limits_{\substack{j=1\\j\neq k}}^{K} \mathbf{p}_{j}^{H}\hat{\mathbf{h}}_{k}\hat{\mathbf{h}}{}^{H}_{k}\mathbf{p}_{j}+\!\!\sum\limits_{j\in \mathcal{I}}{\mathbf{p}_{j}^{H}\mathbf{\Phi}_k\mathbf{p}_{j}}+\sigma _{{n}}^2}\vvast).
\label{Rk2_2}
\end{align}
%
% We combine precoding vector $\mathbf{p}=[\mathbf{p}_{1}^{H},\dots,\mathbf{p}_{K}^{H},\mathbf{p}_{c}^{H}]^{H} \in \mathbb{C}^{N_{t}(K+1) \times 1}$. 
%The transmission power constraint can be expressed $\norm{\mathbf{p}}^2=P_t$.
By using a combined precoding vector $\mathbf{p}=[\mathbf{p}_{1}^{H},\dots,\mathbf{p}_{K}^{H},\mathbf{p}_{c}^{H}]^{H} \in \mathbb{C}^{N_{t}(K+1) \times 1}$, the numerator term in (\ref{Rck1_2}) can be expressed by 
\begin{align}{\sum\limits_{\substack{j\in \mathcal{I}}}{\mathbf{p}_{j}^{H}(\hat{\mathbf{h}}_{k}\hat{\mathbf{h}}{}^{H}_{k}+\mathbf{\Phi}_k)\mathbf{p}_{j}}+\sigma _{{n}}^2}=\mathbf{p}^{H}\mathbf{A}_{k}\mathbf{p},
\end{align}
where $\mathbf{A}_{k}\in \mathbb{C}^{N_{t} (K+1) \times N_{t} (K+1)}$ is a block diagonal and positive definite matrix defined by
\begin{align}
\boldsymbol{\mathrm{A}}_{k}=
\begin{bmatrix}
    \hat{\mathbf{h}}_{k}\hat{\mathbf{h}}{}^{H}_{k}& 0 &\cdots  & 0 \\
    0 & \hat{\mathbf{h}}_{k}\hat{\mathbf{h}}{}^{H}_{k} & \cdots  &0 \\
    \vdots & \vdots  &\ddots & \vdots \\
    0 &0 & \cdots  & \hat{\mathbf{h}}_{k}\hat{\mathbf{h}}{}^{H}_{k}
\end{bmatrix}\nonumber\\
+\left(\frac{\sigma _{{n}}^2}{P_{t}}+\sigma _{\mathrm{e},k}^2\right)\mathbf{I}_{N_t(K+1)}
\label{Ak}
\end{align}
under the assumption that all transmission power is used, i.e., $\norm{\mathbf{p}}^{2}=P_t$, and $\mathbf{\Phi}_k=\sigma _{\mathrm{e},k}^2\mathbf{I}$.
We apply the same approach to the denominator and the numerator terms of  (\ref{Rck1_2}), (\ref{Rk2_2}). Each term can be rewritten as:
\begin{align}
%B
{\sum\limits_{j=1}^{M} {\mathbf{p}_{j}^{H}\hat{\mathbf{h}}_{k}\hat{\mathbf{h}}{}^{H}_{k}\mathbf{p}_{j}}+\sum\limits_{\substack{j\in \mathcal{I}}}{\mathbf{p}_{j}^{H}\mathbf{\Phi}_k\mathbf{p}_{j}}+ \sigma _{{n}}^2}=\mathbf{p}^{H}\mathbf{B}_{k}\mathbf{p},\\
{\sum\limits_{\substack{j=1\\j\neq k}}^{M} \mathbf{p}_{j}^{H}\hat{\mathbf{h}}_{k}\hat{\mathbf{h}}{}^{H}_{k}\mathbf{p}_{j}+\!\!\sum\limits_{j\in \mathcal{I}}{\mathbf{p}_{j}^{H}\mathbf{\Phi}_k\mathbf{p}_{j}}+\sigma _{{n}}^2}=\mathbf{p}^{H}\mathbf{D}_{k}\mathbf{p},
\end{align}
where the positive definite matrix $\mathbf{B}_{k},\mathbf{D}_{k}$ are defined as
\begin{align}
\boldsymbol{\mathrm{B}}_{k}=\boldsymbol{\mathrm{A}}_{k}-
\begin{bmatrix}
    0 &  \cdots & \cdots & 0 \\
    \vdots & \ddots  & \ddots & \vdots \\
    0 &  \cdots & 0 &0\\
    0 &  \cdots  & 0 & \hat{\mathbf{h}}_{k}\hat{\mathbf{h}}{}^{H}_{k}
\end{bmatrix},
\label{Bk}
\end{align}
\begin{align}
\boldsymbol{\mathrm{D}}_{k}=\boldsymbol{\mathrm{B}}_{k}-
\begin{bmatrix}
{0} & \cdots  & 0 & \cdots  & 0 \\
 \vdots& \ddots &\vdots & \ddots &\vdots \\
0 & \cdots  & 
{\hat{\mathbf{h}}_{k}\hat{\mathbf{h}}{}^{H}_{k}} 
& \cdots  & 0  \\
 \vdots & \ddots  &  \vdots & \ddots &  \vdots \\
 0 & \cdots  & 0 & \cdots  & {0} 
\end{bmatrix}.
\label{Dk}
\end{align}
The matrix in the second term of (\ref{Dk}) is a block diagonal matrix formulated by diag($\bm{0},\dots,\hat{\mathbf{h}}_{k}\hat{\mathbf{h}}{}^{H}_{k},\dots,\bm{0}$)
%whose sub-block size is $N_t\times N_t$
in which $\hat{\mathbf{h}}_{k}\hat{\mathbf{h}}{}^{H}_{k}$ is a $k$th diagonal sub-block.
As the result, the achievable rates are simplified to 
\begin{align}
R_{c,k}=\log_{2}\left(\frac{\mathbf{p}^{H}\mathbf{A}_{k}\mathbf{p}}{\mathbf{p}^{H}\mathbf{B}_{k}\mathbf{p}}\right),~R_{k}=\log_{2}\left(\frac{\mathbf{p}^{H}\mathbf{B}_{k}\mathbf{p}}{\mathbf{p}^{H}\mathbf{D}_{k}\mathbf{p}}\right),\label{RR3}
\end{align}
and the objective function of the optimization problem $\bm{\mathsf{(P1)}}$ can also be simplified to
\begin{align}
f(\mathbf{p})=\sum\limits_{k=1}^{K}\log_{2}\left(\frac{\mathbf{p}^{H}\mathbf{B}_{k}\mathbf{p}}{\mathbf{p}^{H}\mathbf{D}_{k}\mathbf{p}}\right)+\min_{j}\log_{2}\left(\frac{\mathbf{p}^{H}\mathbf{A}_{j}\mathbf{p}}{\mathbf{p}^{H}\mathbf{B}_{j}\mathbf{p}}\right).
\end{align}
In this case, because $f(\mathbf{p})=f(\alpha \mathbf{p})$ with non-zero parameter $\alpha$, the power constraint ($\ref{P1:const2}$) can be ignored. Each rate is written as the difference of concave functions, e.g. $R_k=\log_2(\mathbf{p}^{H}\mathbf{B}_{k}\mathbf{p})-\log_2(\mathbf{p}^{H}\mathbf{D}_{k}\mathbf{p})$, which is a non-convex function. Thus, finding the optimal solution is difficult due to the non-convexity of the objective function. 
In order to solve the non-convex problem, we propose an algorithm based on alternating optimization.
%The proposed algorithm combining the two techniques used for non-convex problem. One is semidefinite relaxation (SDR) which obtains a solution that is close to the optimal solution of non-convex quadratically constrained quadratic program (QCQP) \cite{SDR}. The other is concave-convex procedure (CCCP) which guarantees local optimal solution of difference of convex  functions (DC) problem \cite{MM_algorithm}.
%
%
\subsection{Proposed Optimization Algorithm}
Before finding a solution, to transform the problem, we derive upper and lower bounds of the denominator and numerator in (\ref{RR3}) by using auxiliary variables $a_{k},b_{k},c_{k},d_{k}$:
\begin{align}
\mathbf{p}^{H}\mathbf{A}_{k}\mathbf{p}\geq e^{a_{k}},
\,\mathbf{p}^{H}\mathbf{B}_{k}\mathbf{p}\geq e^{c_{k}},
\label{AC:bound}
\end{align}
\begin{align}
\mathbf{p}^{H}\mathbf{B}_{k}\mathbf{p}\leq e^{b_{k}},
\,\mathbf{p}^{H}\mathbf{D}_{k}\mathbf{p}\leq e^{d_{k}}.
\label{BD:bound}
\end{align}
Using these bounds, we can induce the lower bound of objective function $f(\mathbf{p})$ by 
\begin{align}
f(\mathbf{p})&\geq \frac{1}{\ln{2}}\left[\sum\limits_{k=1}^{K}\ln \left(\frac{e^{c_k}}{e^{d_k}}\right)+\min_{j}\ln\left(\frac{e^{a_j}}{e^{b_j}}\right)\right]\nonumber
\\
&=\frac{1}{\ln{2}}\left[\sum\limits_{k=1}^{K}\left(c_{k}-d_{k}\right)+\min_{j}(a_{j}-b_{j})\right].
\label{f:bound}
\end{align}
Finally, by adding one more slack variable $l_{c}$ for term of $\min_{j}(a_{j}-b_{j})$, % we transform problem $\mathsf{(P2)}$ to $\mathsf{(P3)}$.
the optimization problem $\bm{\mathsf{(P1)}}$ can be transformed to $\bm{\mathsf{(P2)}}$ denoted as
\begin{maxi!}|s|[0]
	{\substack{\mathbf{a},\mathbf{b},\mathbf{c},\mathbf{d}\\\mathbf{p},l_{c}}}{\sum\limits_{k=1}^{K}(c_{k}-d_{k})+l_{c}}
	{\label{P3}}{\bm{\mathsf{(P2):}}\nonumber}
	\addConstraint{a_{k}-b_{k}\geq l_{c}, k=1,\dots,K}
	\addConstraint{(\ref{AC:bound}), (\ref{BD:bound})},\nonumber
%	\addConstraint{p_n}{\geq 0, \, \forall n \in \mathcal{N}, \label{P1:const1}}
\end{maxi!}
where $\mathbf{a}\delequal[a_{1},\dots,a_{K}]$, $\mathbf{b}\delequal[b_{1},\dots,b_{K}]$, $\mathbf{c}\delequal[c_{1},\dots,c_{K}]$, and $\mathbf{d}\delequal[d_{1},\dots,d_{K}]$.
The problem is still non-convex, since the constraints (\ref{AC:bound}), (\ref{BD:bound}) are non-convex.
For constraint (\ref{AC:bound}), we apply SDR technique which obtains a solution that is close to the optimal solution in non-convex quadratically constrained quadratic program (QCQP) \cite{SDR}.
SDR converts a non-convex problem into a convex problem by removing the rank one constraint which causes non-convexity.
To apply SDR, we transform the quadratic term of (\ref{AC:bound}) as 
\begin{align}
\mathbf{p}^{H}\mathbf{A}_{k}\mathbf{p}=\mathrm{tr}(\mathbf{p}^{H}\mathbf{A}_{k}\mathbf{p})=\mathrm{tr}(\mathbf{A}_{k}\mathbf{p}\mathbf{p}^{H})=\mathrm{tr}(\mathbf{A}_{k}\mathbf{X})
\end{align}
by converting $\mathbf{p}\mathbf{p}^{H}$ to $\mathbf{X}$ with constraints $\mathbf{X}\succeq 0$ and $\mathrm{rank}(\mathbf{X})=1$. Then the constraint (\ref{AC:bound}) becomes convex when the rank constraint is removed.
Generally, an optimal solution of the relaxed problem may not satisfy the rank constraint, which implies that an additional process is required to construct a genuine solution that satisfies the rank constraint. 
%an optimal objective value of the relaxed problem serves as an upper bound to that of the prior problem and 
%an optimal solution of relaxed problem does not satisfy the rank constraint
%
This issue will be tackled after the algorithm is described.
%Thus, when using SDR, a obtained solution does not satisfy the rank constraint and additional steps are necessary for satisfying the rank constraint. This problem is treated at the end of the algorithm.
%
\begin{algorithm}[t]
	\caption{Alternating Optimization based on SDR and CCCP}
	\label{Algo}
	\begin{algorithmic}[1]
	\STATE Initialize $b_{k}^{(0)}, d_{k}^{(0)}$ and set s=0 and $\epsilon$ to a small value\\
    \STATE \algorithmicrepeat
    \STATE $s\leftarrow s+1$ 
	\STATE Given  $b_{k}^{(s-1)}, d_{k}^{(s-1)}$ , solve the problem $\mathsf{(P3)}$ and obtain optimal $\mathbf{X}^*,a_{k}^*,b_{k}^*,c_{k}^*,d_{k}^*,l_{c}^*$
	\STATE Update $b_{k}^{(s)}\leftarrow b_{k}^*$, $d_{k}^{(s)}\leftarrow d_{k}^*$
	\STATE \algorithmicuntil  \,\, convergence of $b_{k}^{(s)}, d_{k}^{(s)}$
	\STATE Decomposition $\mathbf{X}^*=\mathbf{U \Sigma U}^{H}$
	\STATE Generate enough random vectors $\mathbf{r}\sim \mathcal{CN} (0,\mathbf{I}_{N_{t} (K+1)})$
	\STATE Choose the best $\mathbf{\bar{r}}=\mathbf{U \Sigma}^{1/2}\mathbf{r}$ as a solution $\mathbf{p}^*$
	\end{algorithmic}
\end{algorithm}

The constraint functions in  (\ref{BD:bound}) are difference of convex (DC) functions that are generally non-convex. 
For this problem, we approximate the DC function to a convex function by the CCCP method, which guarantees a local optimal solution of the DC problem \cite{MM_algorithm}. 
For such approximation, we linearize the exponential term in (\ref{BD:bound}), which is concave, by using the first-order Taylor series approximation.
%To approximate, we linearize the exponential term that is concave in (\ref{BD:bound}) by using the first-order Taylor series approximation.
%
Finally, constraints at the $s$th iteration can be denoted by  
\begin{align}
\mathrm{tr}(\mathbf{A}_{k}\mathbf{X})\geq e^{a_{k}},
~\mathrm{tr}(\mathbf{B}_{k}\mathbf{X})\geq e^{c_{k}},
\label{AC:Relx}
\end{align}
\begin{align}
\mathrm{tr}(\mathbf{B}_{k}\mathbf{X})\leq e^{b_{k}^{(s-1)}}(b_{k}-b_{k}^{(s-1)}+1),
\label{B:Relx}
\end{align}
\begin{align}
\mathrm{tr}(\mathbf{D}_{k}\mathbf{X})\leq e^{d_{k}^{(s-1)}}(d_{k}-d_{k}^{(s-1)}+1).
\label{D:Relx}
\end{align}
%where $b_{k}^{(s-1)}$, $d_{k}^{(s-1)}$ is  
The sub-problem at $s$th iteration $\bm{\mathsf{(P3)}}$ is a convex problem expressed by 
\begin{maxi!}|s|[0]
	{\substack{\mathbf{a},\mathbf{b},\mathbf{c},\mathbf{d}\\\mathbf{p},l_{c}}}{\sum\limits_{k=1}^{M}(c_{k}-d_{k})+l_{c}}
	{\label{P3}}{\bm{\mathsf{(P3):}} \nonumber}
	\addConstraint{a_{k}-b_{k}\geq l_{c}, k=1,\dots,K}
	\addConstraint{\mathbf{X}\succeq 0}
	\addConstraint{(\ref{AC:Relx}), (\ref{B:Relx}), (\ref{D:Relx})}.\nonumber
%	\addConstraint{p_n}{\geq 0, \, \forall n \in \mathcal{N}, \label{P1:const1}}
\end{maxi!}
%
%
%The solution of the sub-problem can be reached to a local optimal solution after enough iterations.
The value of $b_{k}^{(s-1)}$ and $d_{k}^{(s-1)}$ are updated by solving the sub-problem $\bm{\mathsf{(P3)}}$ and a local optimal solution is obtained with a sufficient number of iterations. The detailed process is expressed in Algorithm 1. The convex problem can be solved using CVX toolbox \cite{CVX}.

It is noted that a obtained solution $\mathbf{X}^*$ does not satisfy the constraint $\mathrm{rank}(\mathbf{X})=1$. Therefore, we refine $\mathbf{X}^*$ to  satisfy the rank constraint.
%There are one more step to find solution of precoding vector $\mathbf{p}^*$, because we just find $\mathbf{X}^*$ not $\mathbf{p}^*$. Rank one approximation is needed to satisfy rank constraint $\mathrm{rank}(\mathbf{X})=1$ that is removed for relaxing the constraints.
%Because of matrix $\mathbf{X}^*$ is a hermitian positive semidefinite matrix, the matrix $\mathbf{X}^*$ can decomposed as
%
Since $\mathbf X^{*} $ is a Hermitian positive semidefinite matrix, it can be decomposed in the form $\mathbf{X}^*=\mathbf{U \Sigma U}^{H}$ by the singular value decomposition. Then, we generate sufficient number of random vectors $\mathbf{r}\in \mathbb{C}^{N_{t}(K+1)\times1} $ and obtain $\mathbf{\bar{r}}=\mathbf{U \Sigma}^{1/2}\mathbf{r}$. Finally, we choose the best $\mathbf{\bar{r}}$ maximizing the objective function $f(\mathbf{p})$ as a final solution $\mathbf{p}^*$.
It has been shown that SDR with sufficiently large number of random vectors guarantees a solution close to the optimal solution \cite{SDR}.
%In these randomization approach, the obtained solution $\mathbf{p}^*$ is closed to the optimal solution with significant amount of random vectors \cite{SDR}. 
%
Overall step of the proposed algorithm is described in Algorithm \ref{Algo}.
\begin{figure}[t]
\includegraphics[width=1\linewidth]{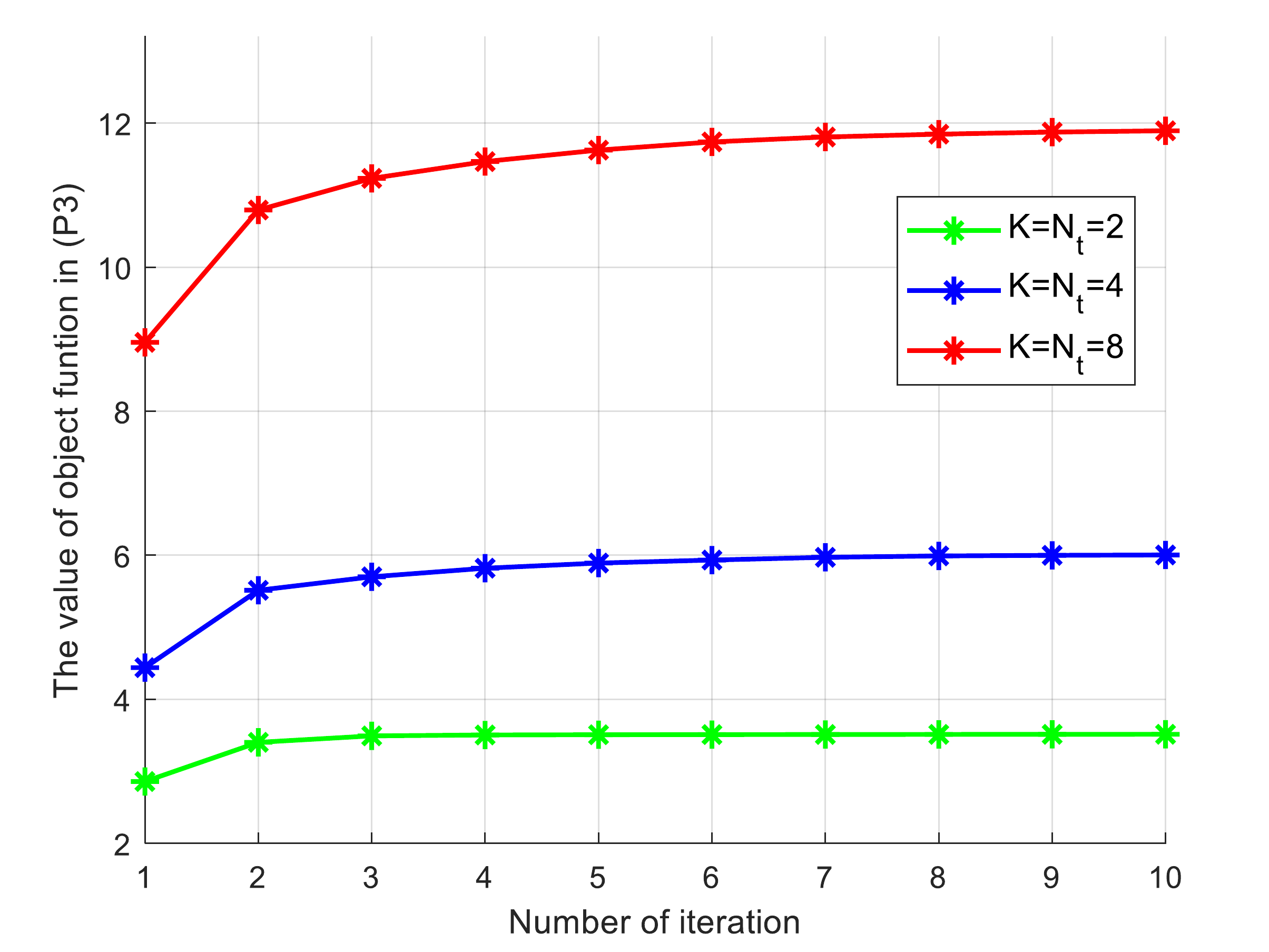}
\centering
\caption{Convergence of proposed algorithm according to the number of transmitter antennas and users}
\label{Converge}
\end{figure}

$\mathbf{Remark~(Convergence):}$ In Algorithm 1, we randomly initialize the values of $b_{k}^{(0)}$ and $d_{k}^{(0)}$.
In Fig. \ref{Converge}, it is numerically confirmed that the proposed algorithm converges to finite point as the number of iterations increases when SNR$=20$dB, $\sigma_{\mathrm{e},k}^2=0.1$. The value of objective function gradually increased without fluctuation.
We can observe that 
%speed of convergence dependent on number of user and transmitter antenna. A
as the number of transmitter antennas and users increases, the number of iterations required for convergence increases.
\section{simulation results}
 We assume $\mathbf{h}_{k}$ has i.i.d complex Gaussian distribution with zero mean and unit variance, i.e. $\mathbf{h}_{k}\sim \mathcal{CN}(0,1)$. The variance of AWGN is fixed as $\sigma_{n}^2=1$. The channel error is also distributed by complex Gaussian distribution, i.e., $\mathbf{e}_{k}\sim \mathcal{CN}(0,\sigma_{\mathrm{e},k}^2 \mathbf{I})$. The estimated channel $\hat{\mathbf{h}_{k}}$ is independent from the channel error and has complex Gaussian distribution with zero mean and variance $\sigma_{\hat{\mathbf{h}}_{k}}^2  =1-\sigma_{\mathrm{e},k}^2 \mathbf{I}$, i.e. $\hat{\mathbf{h}}_{k}\sim \mathcal{CN}(0,1-\sigma_{\mathrm{e},k}^2 \mathbf{I})$. We also consider that all channel error has same covariance matrix $\sigma_{\mathrm{e},k}^2\mathbf{I}=\sigma_{\mathrm{e}}^2\mathbf{I}$.

\subsection{Comparison with Conventional Multiple Access Techniques}
 In this section, we consider a 2-user scenario and provide simulation results to compare with existing multiple access strategies: SDMA, NOMA, and OMA. It has been shown that these conventional strategies are special cases of RSMA in a 2-user scenario \cite{RS:uni}. As shown in TABLE \ref{tab1}, RSMA can boil down to conventional strategies depending on the power levels allocated to the streams. 
% work as the conventional strategies according to power allocation. 
When user-1 has a stronger channel than user-2, the private stream of user-2 should be turned off, resulting in NOMA. 
Regardless of the number of users, RSMA works as SDMA when the common stream is turned off. Thus, the optimal power allocation and precoding vectors in NOMA and SDMA can be carried out by modifying Algorithm 1.
Specifically, the precoding vector of the private stream of user-2 is set to zero vector in case of NOMA, while that of the common stream is set to zero vector in case of SDMA.
Also, RSMA can be reduced to OMA by allocating the total transmit power to one private stream within a time slot.
For OMA, we apply maximum ratio transmission (MRT) to precoding vector and assume that the same time resource is allocated to user-1 and user-2 for fairness.
\begin{figure}[t]
\includegraphics[width=1.03\linewidth]{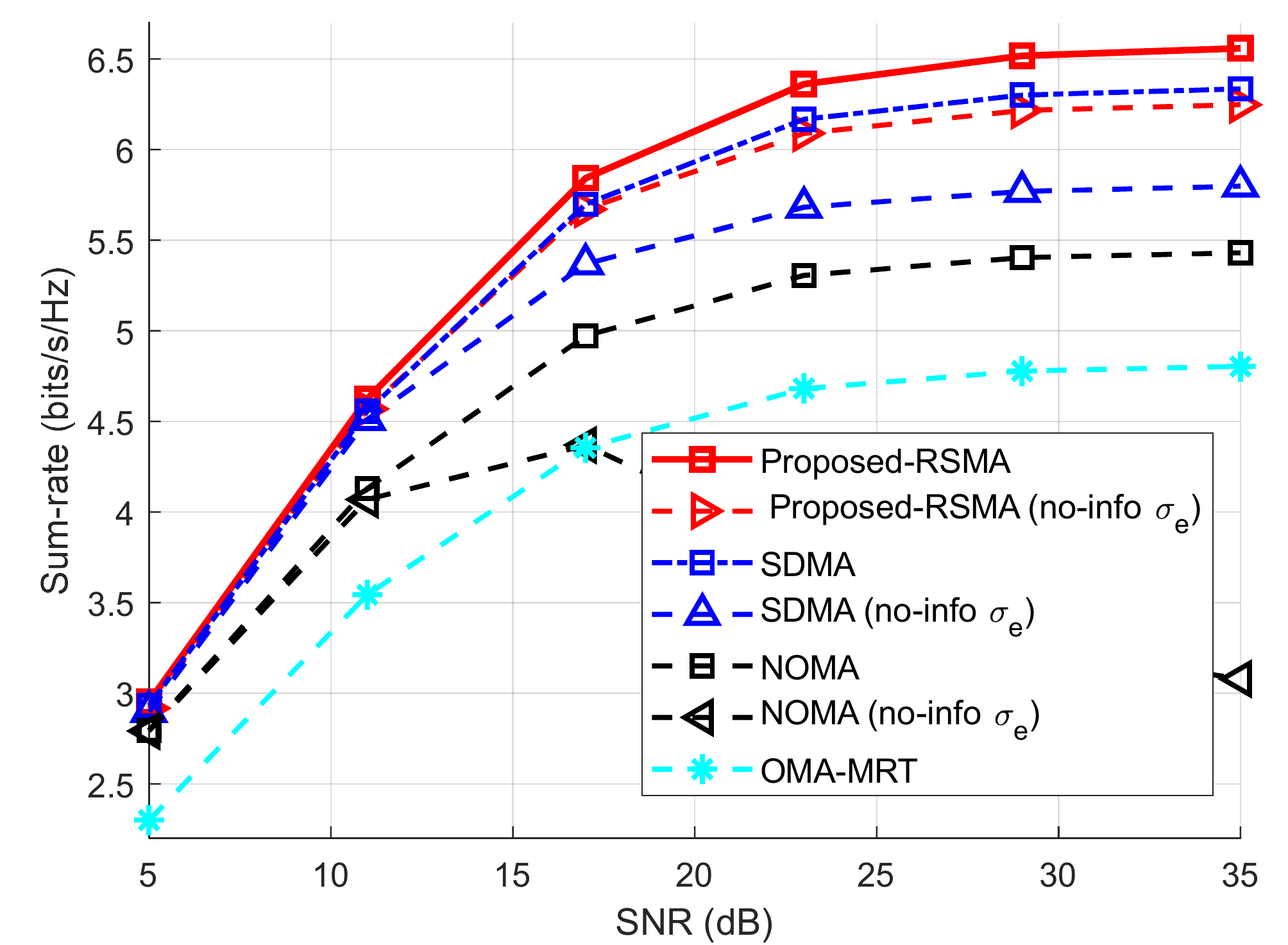}
\centering
\caption{ Sum-rate comparison between RSMA and convectional schemes, where $\sigma_{\mathrm{e}}^2=0.05$, $K=2$, and $N_t=2$}% versus SNR}% when $\sigma_{\mathrm{e}}^2=0.05$}
\label{SNR1}
\end{figure}
%   
%
%
%\newcolumntype{a}{>{\columncolor{Gray}}c}
\begin{table}[t]
\caption{power assigned to each stream according to different multiple access schemes}
\begin{center}
\begin{tabular}{|c|c|c|c|}\hline
\cellcolor{Gray}$\mathbf{Multiple~Access}$& ~~~~$s_1$~~~~&~~~~$s_2$~~~~& ~~~~$s_c$~~~~\\
\hhline{|=|=|=|=|}
\cellcolor{Gray}$\mathbf{RSMA}$&$P_1$& $P_2$& $P_c$ \\
\hline
\cellcolor{Gray}~~$\mathbf{NOMA}$~~&$P_1$& 0& $P_c$ \\
\hline
\cellcolor{Gray}$\mathbf{SDMA}$&$P_1$& $P_2$& 0 \\
\hline
\cellcolor{Gray}$\mathbf{OMA}$&$P_1$&0& 0 \\
\hline
\end{tabular}
\label{tab1}
\end{center}
\end{table}

In order to confirm the usefulness of the channel error information, a scenario in which the BS has no information about channel error, labeled as no-info $\sigma_e$, is further considered with the schemes described above. 
In other words, the BS optimizes the precoding vectors by considering $\hat{\mathbf{h}}_{k}$ as the perfect channel estimate.
Note that in OMA, since the precoding vector and the power allocation are fixed, the sum-rate performance is not changed depending on whether or not there is no information about channel error in the BS.
%
%As an exception, in OMA, it is not considered the scenario with the absence of channel error information because the precoding vector is precoded by MRT relying on the estimated channel regardless of the channel error.
%

We illustrate the achievable sum-rate of the proposed method according to SNR when the BS has two antennas and  $\sigma_{\mathrm{e}}^2=0.05$.
As shown in Fig. \ref{SNR1}, RSMA has a better performance than other multiple accesses.
The gap in sum-rate between RSMA and the other schemes is notable in high SNR.
It is worth pointing out that these benefits come from the flexibility of RSMA that generalizes and bridges the conventional schemes for 2-user scenario.
Compared to SDMA, the use of the common stream for RSMA, offers more design flexibility in jointly optimizing its precoding vector and power allocation.
%since RS has the common part stream, there are more opportunities for design the precoding vectors and power allocation in RS, which generates gains in the sum-rate.
%
%
%RS has gains over SDMA with respect to the sum-rate.
%These results show that the flexibility of RS, generalizing the conventional schemes, derives benefits with respect to the sum-rate and robustness over imperfect CSI compared to the conventional schemes.
%This result shows that flexibility of RS, sharing the power to the common stream and the private stream, is more efficient to increasing sum-rate than sharing power to some specific streams by turning-off the others.
%
Under perfect CSIR, there is no increment of interference when a desired signal transmit power is increased.
However, under imperfect CSIR, the interference from the channel error is also increased as the desired signal transmit power is increased.
%Since both the denominator and numerator terms of GMI can be expressed as linear functions of signal power, GMI is bounded by the signal power.
Thus, GMI and the sum-rate are saturated at high SNR.
%GMI no longer increases when the desired signal transmit power is high enough and the sum-rate is saturated at high SNR. 
%allocating total power to specific the private or common steam when transmission power is large enough.
%
Furthermore, it is described in Fig. \ref{SNR1} that the performance is degraded in the absence of information about channel error and RSMA is less sensitive to the knowledge on channel errors than the other schemes. 
\begin{figure}[t]
\includegraphics[width=0.97\linewidth]{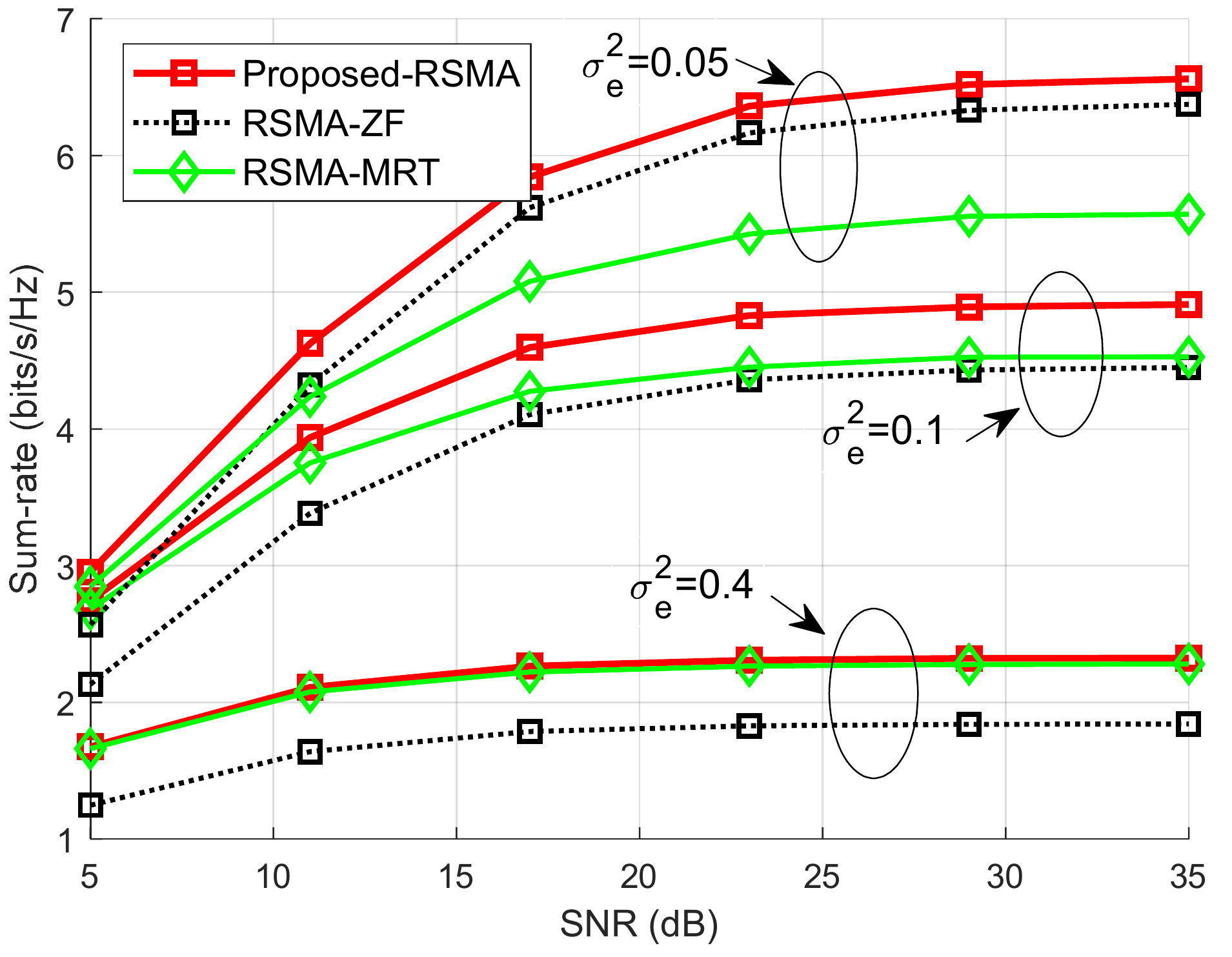}
\centering
\caption{Sum-rates of RSMA achieved by proposed and fixed precoding vectors according to SNR and $\sigma_e^2$, where $K=2$ and $N_t=2$}
\label{SNR2}
\end{figure}
\begin{table}[t]
\caption{Precoding vector of private stream with ZF and MRT}
\begin{center}
%\tabulinesep=0.8mm
\begin{tabular}{|c|c|c|c|}
\hline
 \cellcolor{Gray}$\mathbf{Precoding}$& Direction of precoding vector for private stream\\
\hhline{|=|=|}
\cellcolor{Gray}&{}\\[-0.7em]
\cellcolor{Gray}$\mathbf{ZF}$& \begin{tabular}{@{}c@{}}$\mathbf{v}_{k}=\mathbf{w}_{k}/\norm{\mathbf{w}_{k}}$ where $ \mathbf{W}=\hat{\mathbf{H}}^{H}(\hat{\mathbf{H}}\hat{\mathbf{H}}^{H})^{-1},$ \\   $\mathbf{W}=[\mathbf{w}_{1},\cdot\cdot\cdot,\mathbf{w}_{K}]$, and $\hat{\mathbf{H}}=[\hat{\mathbf{h}}_1,\cdot\cdot\cdot,\hat{\mathbf{h}}_K]^H$\end{tabular}
\\
\cellcolor{Gray}&{}\\[-0.7em]
\hline
\cellcolor{Gray}&{}\\[-0.7em]
\cellcolor{Gray}$\mathbf{MRT}$&$\mathbf{v}_{k}=\hat{\mathbf{h}}_{k}/||{\hat{\mathbf{h}}}_{k}||$\\[3pt]

\hline
\end{tabular}
\label{tab2}
\end{center}
\end{table}
\subsection{Impact of Proposed Precoding in RSMA}
 %In respect to precoding, comparison with zero-forcing (ZF) and maximum ratio transmission (MRT) are analyzed.
In this section, performances of the precoding vectors are analyzed by comparing simulation results of RSMA with the proposed precoding vectors and existing fixed precoding vectors, zero-forcing (ZF) and MRT.
  %RS with existing precoding schemes, zero-forcing (ZF) and MRT, is investigated for a performance analysis of the precoding vector optimized by the proposed algorithm.
  %zero-forcing (ZF) and MRT are analyzed and we numerically show the impact of optimizing percoding vector of the private stream.
ZF beamforming aims to remove the interference by nulling, i.e. $|\hat{\mathbf{h}}{}^{H}_{k}\mathbf{p}_{j}|=0,~j \neq k$ \cite{ZF-BF} and MRT refers to precoding the stream in the same direction with the channel vector.
%
%ZF and MRT are applied to the precoding vectors of the private steams and determine the direction of the precoding vector based on the estimated channel.
ZF and MRT are applied to the precoding vectors of the private streams. As shown in TABLE \ref{tab2},  ZF and MRT determine the direction of the precoding vector for the private stream based on the estimated channel.
%
%
  %We apply ZF and MRT to the private signal of RS. 
  %As shown in TABLE \ref{tab2}, we can generate precoding vector using ZF and MRT based on the estimated channel. 
As a result, the direction of the precoding vectors for the private stream, $\mathbf{v}_k=\mathbf{p}_k/\norm{\mathbf{p}_{k}},~k=1,\dots,K$, are fixed. Thus, power allocation, $P_k=\norm{\mathbf{p}_{k}}$, and precoding vector for the common stream ${\mathbf{p}_{c}}$ should be optimized. 
%
 %When using ZF, interference between the private signals is removed, i.e. $|\hat{\mathbf{h}}{}^{H}_{k}\mathbf{p}_{j}|=0,~j \neq k$.
% We optimize sum-rate of RS with ZF using same approach for proposed RS.
 We optimize the sum-rate of RSMA-ZF/MRT with the similar approach of the proposed RSMA by formulating an optimization problem for $P_k$ and ${\mathbf{p}_{c}}$.
 %However, unlike proposed RS, rate is not simplified as function of $\mathbf{p}$, power constraint should not be removed.
%In the optimization process, a difference between RS with ZF/MRT and proposed RS is that SDR is used for quadratic term of precoding vector for common signal. %CCCP still used for exponential terms. 

 %In RS combined with MRT, precoding direction of private signal is determine by channel direction, i.e. $\mathbf{v}_{k}=\hat{\mathbf{h}_{k}}/||{\hat{\mathbf{h}_{k}}}||$.
 %Unlike rate in perfect CSI, there is one more noise term in imperfect CSI. 
 % In Fig.\ref{SNR1}, it also demonstrated that the existence of channel error information considerably effect to performance. 
%
Fig. \ref{SNR2} demonstrates the changes in the performance of RSMA with respect to channel error covariance $\sigma_e^2$ and SNR, where $K=2$ and $N_t=2$.
The results confirm that the rate under imperfect CSI has different characteristics from that obtained under perfect CSI.
Under perfect CSI, it is well known that MRT is near optimal in low SNR and ZF is asymptotically optimal in high SNR \cite{ZF_MRT}.
However, the imperfect knowledge of channel brings out the residual interference caused by the inaccurate operations of precoding (at the transmitter) as well as coherent detection including SIC (at the receiver), which corresponds to low SNR scenarios. 
When the variance of channel error, $\sigma_e^2$, is dominant, RSMA tends to operate as in low SNR under perfect CSI so that the optimized beamformers approximate to MRT.
On the other hand, when channel error variance is very small, e.g. $\sigma_e^2\approx0$, ZF can provide a better sum-rate perforamnce than that of MRT due to negligible residual interference at high SNR regime.
As shown in Fig. \ref{SNR2}, RSMA with the proposed precoding vector performs stricly better than RSMA with ZF and MRT precoders regardless of the variance of channel error over the entire range of SNR.
%
%Consequently, due to the influence of the channel error, selecting a precoding vector as MRT or ZF according to SNR is a very naive method and the precoding vector needs to be optimized depending on the situation.
%
%It is observed that the proposed method optimizing precoding vector of the private stream has a better sum-rate performance than fixed precoding schemes regardless of the SNR and error covariance.

  %
\section{Conclusion}
In this paper, we have studied a robust design of RSMA when perfect CSI is not available at both transmitter and receiver.
To tackle the sum-rate maximization problem turned out to be non-convex, the proposed algorithm has utilized the SDR and CCCP in jointly optimizing the precoding vectors and power allocation.
%
%Based on the proposed algorithm, the sum-rate has been maximized by jointly optimizing the precoding vectors and power allocation.
%address the sum-rate maximization problem by optimizing the precoding vector in RS approach with imperfect CSIT an CSIR.
%which is combined direction and power
%for the sum-rate maximization. 
%We have formulated the sum-rate maximization problem that has non-convexity.
%
%In simulate results, it has been numerically shown that the improvement of the sum-rate 
The simulations results have numerically shown that the proposed RSMA achieves the enhanced sum-rate performance compared to the conventional multiple access schemes. 
Also, RSMA with joint optimization of power allocation and the precoding vectors provides the sum-rate improvement over RSMA in which the fixed precoding schemes, ZF and MRT, which are applied to the private streams.
%Also, the impact of the precoding vector adapted by proposed algorithm has been shown through the performance comparison with RS with ZF/MRT which fixes the direction of precoding vector.
%the simulation results showed that superiority of the precoding vector optimized with the proposed algorithm compared to the fixed precoding schemes.
%
From these results, it can be seen that the proposed RSMA is a powerful technique in terms of the sum-rate performance under imperfect CSIR and CSIT.

 \section*{Acknowledgment}

This work was supported by the Basic Science Research Programs
under the National Research Foundation of Korea (NRF) through the Ministry
of Science and ICT under Grants NRF-2019R1C1C1006806.

 \bibliographystyle{IEEEtran} % uncomment for IEEE style

%\bibliography{Reference1}\label{refs}

\end{document}